# Microwave heating effect on diamond sample of NV centers

*Wang Zheng* [a, b] *Zhang Jintao* [b, *], *Feng Xiaojuan* [b], *Xing Li* [b]

[a] Tsinghua University, Beijing, China, email address: zheng-wa18@mails.tsinghua.edu.cn
[b] National Institute of Metrology, Beijing, China, email address: fengxj@nim.ac.cn
* Corresponding author. E-mail address: zhangjint@nim.ac.cn

**Diamond samples of defects with negative charged nitrogen-vacancy (NV) centers are promising solid state spin sensors suitable for quantum information processing, high sensitive measurements of magnetic, electric and thermal fields in nanoscale. The diamond defect with a NV center is unique for its robust temperature-dependent zero field splitting $D_{gs}$ of the triplet ground state. This property enables optical readout of electron spin states through manipulation of the ground triplet state using microwave resonance with $D_{gs}$ from 100 K to about 600 K. Thus, prohibiting $D_{gs}$ from unwanted external thermal disturbances is crucial for an accurate measurement using diamond NV sensors. Our observation demonstrates the existence of a prominent microwave heating effect on the diamond samples of NV centers. The effect is inevitable to shift $D_{gs}$ and cause measurement errors. The temperature increment caused by the effect monotonically depends on the power and the duration of microwave irradiation. The effect is obvious with the microwave irradiation in the continuous mode and some pulse sequence modes, but is neglectable for the quantum lock-in XY8-N method.**

The negative charged nitrogen-vacancy (NV) center is a unique defect in diamond that is distinguished for its optical spin polarization and readout in room temperatures [1,2,3,4,5]. This specific property makes NV centers promising as solid state spin sensors suitable for quantum physics studies and information processing [6,7,8,9,10,11], nanoscale measurements of magnetic and electric fields[12,13,14,15,16], temperatures [17,18,19,20,21,22,23,24]. The spin projection split by a zero-field splitting $D_{gs}$ enables the spin sublevels of the triplet ground state detectable by the optically detected magnetic resonance (ODMR) in the continuous wave (CW) mode and the pulse wave sequence mode. In ODMR techniques, a microwave resonances with $D_{gs}$ enables manipulation of the spin state distribution to yield difference in the optical emission intensity upon optical excitation. Thus, the electron spin states are accessible by such manipulation. According to literature [25], the electronic spin Hamiltonian of the ground state spin $\hat{H}_{gs}$ is:

$$\hat{H}_{gs} = \frac{1}{\hbar^2}(D_{gs} + d_\parallel \Pi_z)S_z^2 + \frac{\mu_B}{\hbar}\vec{S}\cdot\overline{g}\cdot\vec{B} - \frac{1}{\hbar^2}d_\perp \Pi_x(S_x^2 - S_y^2) + \frac{1}{\hbar^2}d_\perp \Pi_y(S_xS_y + S_yS_x) \quad (1),$$

where $\hbar$ stands for Planck constant, and $D_{gs}$ is the zero-field splitting that is temperature dependent, $d_\parallel$ and $d_\perp$ denotes the parallel and perpendicular electric dipole parameters, the total effective field $\vec{\Pi} = \vec{\sigma} + \vec{E}$ constitutes of the electric field $\vec{E}$ and the effective strain field $\vec{\sigma}$, $S_x$, $S_y$ and $S_z$ are the x, y, and z components of the spin operator, respectively. $\overline{g}$ is the Lande factor, and $\mu_B$ is the Bohr magneton, $\vec{B}$ is the external magnetic field.

Diamond is a semiconductor of multiple energy bands. A transmitting microwave shall interact with the intrinsic fields in diamond causing thermal energy dissipation in the sample [26]. We name such a formation of heat arising from the dissipation as the microwave heating effect. Thus, a microwave manipulation probably raises a perturbation to $D_{gs}$. Some researchers concerned of the microwave heating effect for readout of the electron spin states of NV centers. Neumann, et al. [27] and Wang, et al. [14] varied powers of microwave in order to find the specific irradiation that causes non-observable change of $D_{gs}$. Toylia, et al. [17] argued that, in the microwave pulse sequence mode, practical microwave irradiation timescale is about three orders lower than the spin coherence times in a measurement. In that case, the heating effect is probably to be omitted. Xie, et al. [28] intended a compensation to the disturbance of heating arising from coplanar waveguide. Scott, et al. [29] pointed out that microwave heating might have a great influence on the experiment of high Rabi frequency. To our knowledge, the microwave heating effect has been yet to be quantitatively and systematically addressed.

Here we demonstrate our quantitative observations on the microwave heating effect on the diamond samples of NV centers. We studied the effect by the typical continuous and pulse sequence microwave irradiation in variant powers. An infrared thermal imager and two thermocouples were applied for measuring thermal images and local temperatures on the surface of the diamond samples. The experimental setup is pictured in Fig. 1a. The diamond sample of NV centers was made by Chemical Vapour Deposition (CVD) synthesis processes with the dimensions in 2.6 mm × 2.6 mm × 0.3 mm from Element Six. The concentration of nitrogen in the sample is less than 1ppm. The diamond sample was glued to a glass-slide in 24 mm×24 mm×0.15 mm. The ensemble was fixed on a gold-plated printed circuit board (PCB). A thin straight wire antenna constituting a copper wire of 60 μm in diameter closely attached to the upper surface of the diamond samples. We prepared two such ensembles, NV1 and NV2. The difference between the two ensembles lies in that, NV1 was installed in an open environment without controlling temperature, and NV2 was accommodated in a temperature-controlled enclosure. The measurements with NV1 were implemented simultaneously by the infrared thermal imager and the thermocouples. The infrared thermal imager is the type FLIR T650sc. The detector is of 640 × 480 pixel. The temperature sensitivity is better than 0.1 K at room temperatures. The focal length is 22 mm. The spatial resolution is 0.41 mrad. The nominal accuracy is ±1 K. The camera can picture in a discrete mode and the continuous mode in frequency of 30 Hz. The thermocouples are type K. They were diagonally attached on the upper surface of the diamond sample shown in Fig. 1b. Their shields are 0.076 mm in outer diameter. Their reference junctions were properly set in an ice thermostat during measurements. The temperatures inside the enclosure accommodating NV2 were controlled within ±0.1 K surrounding the target temperature. To rule out the effect of temperature control on observing the observed microwave heating effect, we adopted the following control scheme where the controller is turned off when the temperature reaches the target temperature, but when the temperature is below the target temperature the controller is turned on. Because the temperature-controlled enclosure hinders the optical path of the infrared thermal imager, the temperature can only be measured with thermocouples.

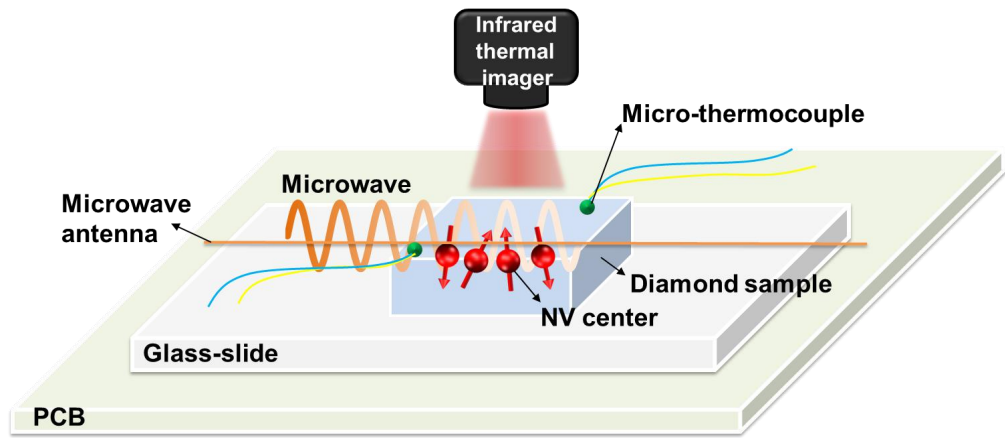

Fig.1 The schematic diagram of the experimental setup

A Vector Signal Generator (Rohde & Schwarz SMIQ04B), tuning from 300kHz to 4.4GHz in the resolution of 0.1 Hz, was used for microwave generation. The synthesized microwaves were transmitted, through a radio frequency (RF) switch (Mini-Circuits ZASWA-2-50DRA+) and an amplifier (Mini-Circuits ZHL-16W-43-S+) of specified range from 1.8 GHz to 4 GHz, to the straight wire antenna. We used the pulse generator (SpinCore PulseBlaster PBESR-PRO-500) specified of the pulse resolution and shortest pulse of 2 ns for generation of pulse signals controlling microwave RF switch. Based on the above devices, we simulated the microwave irradiation of the typical CW, Rabi pulse sequence, spin echo sequence and XY8-10 pulse sequence, and study the microwave heating effect on the ground electron triplet spin state. All the measurements reported here were implemented at the microwave frequency of 2870 MHz, a typical zero-field splitting frequency near room temperature [17].

We first investigated the heating effects of the NV1 sample under continuous microwave irradiation of -13 dBm, -20 dBm, -30 dBm. The duration of each irradiation was 60 s to simulate the general CW-ODMR process. FLIR T650sc was operated in a continuous image mode. A significant microwave heating effect was record in the infrared images. The temperature of the sample is related to the microwave irradiation. We selected the infrared images under -20 dBm microwave irradiation for the demonstration in Fig. 2a. The diamond sample was observed in a thermal equilibrium with the substrate and PCB before the onset of irradiation. After the irradiation begins, the temperature rises rapidly in the diamond sample， while the temperature increase of the substrate is limited, and the temperature rise of PCB is hardly observed. The hottest spots are present in the center of the diamond sample. The real surface temperature of the sample is extracted using a radiation thermometer, depending on accurately knowing the sample surface emissivity at the corresponding wavelength. However, the accurate information on the actual emissivity of the diamond sample is unknown. We applied the instrument's default emissivity when operating the FLIR. Thus, the FLIRT650sc readings are nominal, but these images correctly record the relative thermal tone on the surface of the NV1 diamond sample, reflecting the temperature distribution.

In all measurements, the two thermocouple readings varied within 0.1 K. We took the average of the two thermocouple readings as a quantitative measurement of the sample surface temperature. In Fig.2b, we plotted the time dependence of the temperatures measured under -20 dBm microwave irradiation for NV1 and NV2. Both relations are quite similar. The microwave heating effect is evident in both cases. The effect is stronger for NV2. The maximum rise rate is 3.48 $K \cdot s^{-1}$ for NV2 and 1.73 $K \cdot s^{-1}$ for NV1. The total temperature increment of NV2 is 16.16 K and NV1 is 12.5 K. The samples were rapidly heated during the first 10 s of microwave irradiation, with NV2 and NV1 respectively increasing in temperature of 13.48 K and 9.14 K. The subsequent temperature rise rate slowed significantly. Once the microwave irradiation is turned off, the sample rapidly loses its temperature in the first 10 seconds, and then asymptotically approaches thermal equilibrium. We plot the maximum histograms of the temperature rise at all the microwave irradiation power of NV1 and NV2 in Fig. 2c. The histograms show that the microwave heating effect is monotonically related to the microwave irradiation power. The maximum increment caused by -30dBm irradiation was one order smaller than the maximum increment caused by -20dBm irradiation. Therefore, reducing the irradiation power is more suitable to reduce the microwave heating effect. All the measurements by thermocouples can be found in the supporting information of this letter.

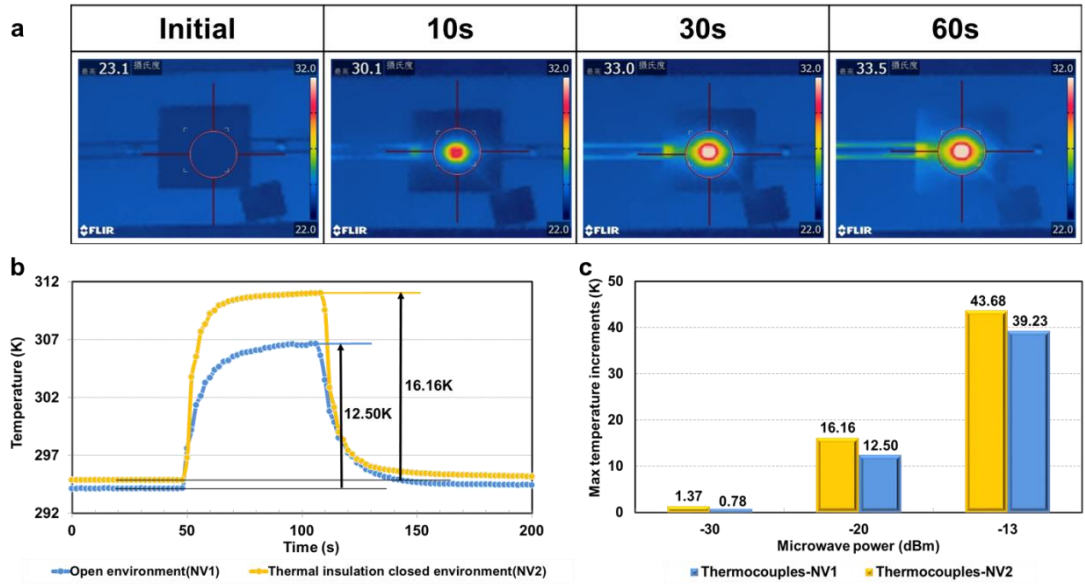

**Fig. 2. Temperature spectra for CW-ODMR. a,** Thermal images for NV1 on the irradiation of -20 dBm. **b**, Temperature spectra for NV1 and NV2 upon the microwave irradiation of -20 dBm. **c**, Maximum histograms of temperature rise at all the microwave irradiation power of NV1 and NV2.

We investigated the microwave heating effect on the cases of pulse microwave sequences. Our study mainly focused on NV2 with -13 dBm and -20 dBm microwave irradiation. First, we investigated the microwave heating effect on the case under microwave irradiation using a general Rabi oscillation. The Rabi pulse sequence is shown in Fig. 3a, a sequence consisting of a series of pulses. The initial pulse was 2 ns wide, and the next pulse was twice the width of the previous pulse. The last pulse was 200 times the width of the initial pulse. The sequence was repeated $10^5$ times for a total duration of approximately 64s. In Fig. 4a, we compared the correlation of temperature via time measured under -13 dBm and -20 dBm microwave irradiation. The correlations are similar to their counterparts under CW irradiation, but the temperature rise is significantly reduced. The maximum rise at -13 dBm and -20 dBm is 2.49 K and 0.64 K, respectively. The temperature rise under higher irradiation is nearly one order larger than that under the lower irradiation. We may reasonably infer that the maximum temperature rise under -30 dBm microwave irradiation may be in the 0.1 K range.

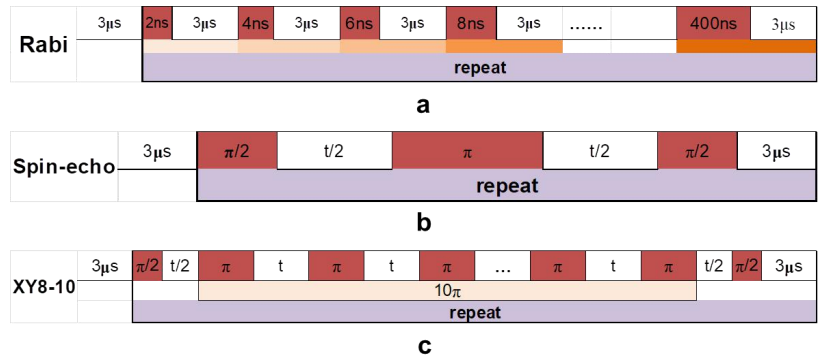

**Fig. 3. Pulse sequences**. The sequence spares 3 μs before conducting an optical readout. **a**, Rabi sequence. The microwave time increased 200 times in 2 ns steps. **b**, Spin echo sequence. The π/2 pulse has a width in 50 ns spaced by the free evolution time (*t*) in 240 μs. **c**, Quantum lock-in XY8-10 sequence. The π/2 pulse has a width of 50 ns spaced by the the free evolution time (*t*) in 60 μs.

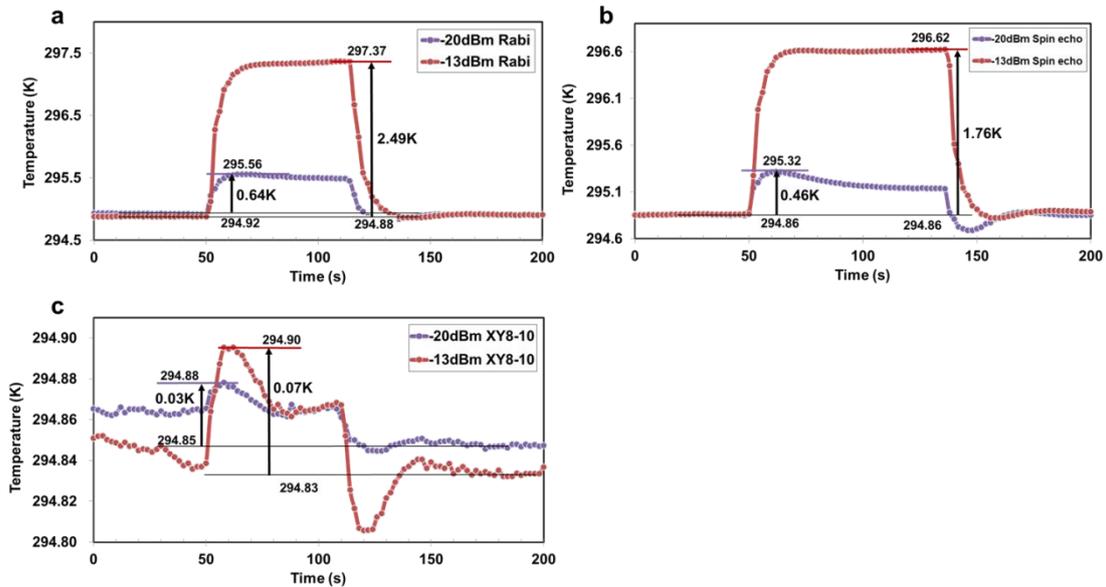

**Fig. 4. Temperature spectra by pulse microwave sequences on the irradiation of -13 dBm and -20 dBm. a**, Rabi sequence. **b**, Spin echo. **c**, The quantum lock-in XY8-10.

Secondly, we studied the case of the general spin echo sequence. As shown in Fig. 3b, the sequence consists of two π/2 pulses and an intermediate π pulse, pulse spaced by the free evolution time. The sequence was repeated $10^5$ times for a total time of approximately 86 s. We compared the temperature time relations for -13 dBm and -20 dBm microwave irradiation in Fig.4b, highly similar to CW irradiation and the Rabi pulse sequence, but the temperature rise is further reduced. The maximum temperature rise is 1.76 K for -13 dBm and 0.46 K for

-20 dBm. The maximum temperature rise for -13 dBm is nearly one order larger than that for -20 dBm. From this, we can infer that the maximum rise in -30 dBm is within the 0.1 K range.

Thirdly, we investigated the case of a quantum lock-in XY8-N sequence, where N was taken to be 10. As shown in Fig. 3c, the sequence consists of two π/2 pulses and 10 intermediate π pulses spaced by the free evolution time. The sequence was repeated $10^5$ times for a total time of approximately 60 s. We compared the temperature time relations for -13 dBm and -20 dBm microwave irradiation in Fig. 4c. The maximum temperature rise is 0.07 K for -13 dBm and 0.03 K for -20 dBm. The temperature rises in both cases are largely reduced with comparing to those for the Rabi oscillation and the spin echo sequence.

In summary, our study demonstrates the existence of prominent microwave heating when reading the electronic ground triplet state of NV centers using ODMR based on continuous microwave, Rabi oscillation, or spin-echo methods. The temperature rise induced by this effect is related to the irradiation power of the microwave and the process of generating the irradiation. For instance, for continuous microwave irradiation, Rabi oscillations, and spin-echo methods, the temperature increment under -20dBm irradiation varies between 16.16K, 0.64K and 0.46K. In contrast, the quantum-locked XY8-N pulse method produces a temperature rise of only 0.03K. Considering that zero-field splitting $D_{gs}$ is temperature dependent, approximately 74 kHz·$K^{-1}$ at room temperature, the microwave heating effect necessarily leads to an alteration for $D_{gs}$, and imposes an external thermal perturbation on measurements of sensors of diamond NV centers using the ODMR procedure. In all observations, the microwave heating effect is negligible for the quantum-locked XY8-N method.

**Acknowledgements** The research has been supported by Fundamental Research Program of National Institute of Metrology, China (No. AKYZD1904-2) and China Postdoctoral Science Foundation (No. 2021M703049).

**Supplementary information**

We present in this supporting information all the pictures of the temperature spectra for NV1 in Fig. S1. The individual microwave irradiation power is in -13 dBm, -14 dBm, -20 dBm and -30 dBm, respectively. The spectra demonstrate quite similar patterns. The maximum temperature increments decrease with decreasing microwave irradiation powers. We present in Fig. S2 the comparison between the temperature spectra for NV1 and NV2. As we stated in the letter, NV1 remarks the case that the diamond sample was set in open environment, and NV2 the case that the diamond sample is set in a thermal insulated enclosure. The spectra demonstrate the similar patterns over all the irradiation levels. The temperature increments with NV2 are constantly larger than those with NV1.

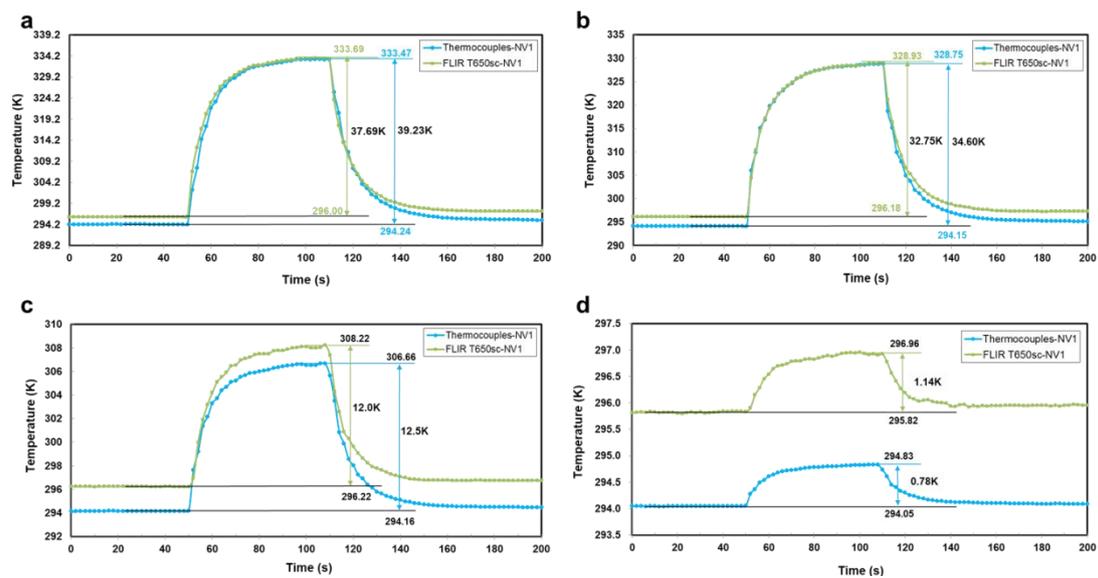

**Fig. S1. Temperature spectra for NV1 obtained by the infrared imager FLIR T650sc and the thermocouples in different microwave power. a,** -13 dBm. **b**, -14 dBm. **c**, -20 dBm. **d**, -30 dBm.

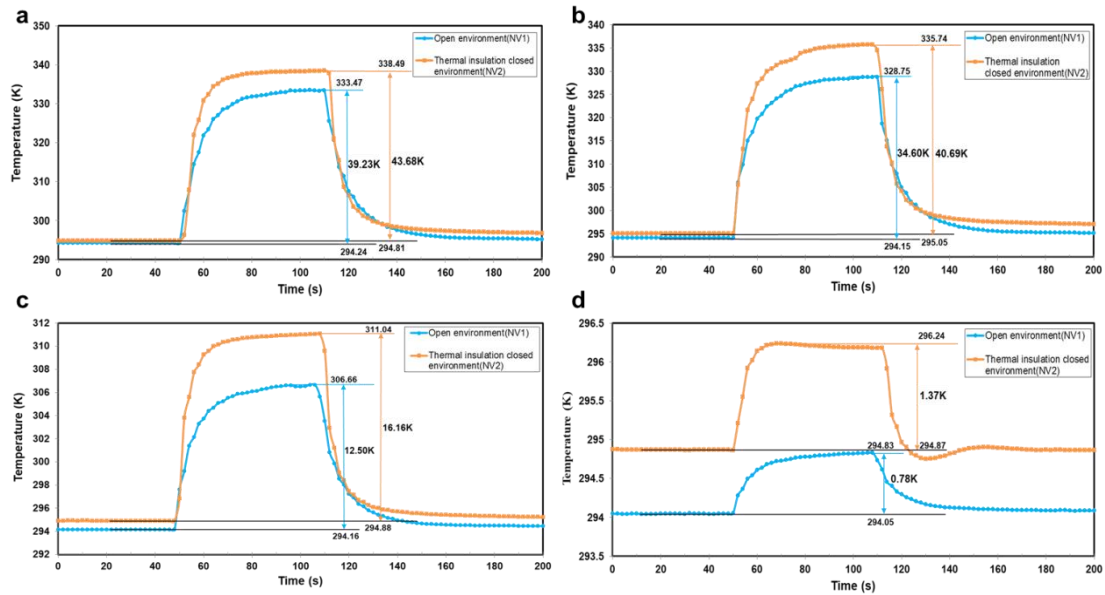

**Fig. S2. Temperature spectra comparison for NV1 and NV2 in different microwave power. a,** -13 dBm. **b**, -14 dBm. **c**, -20 dBm. **d**, -30 dBm.